\documentclass[11pt,amsmath,amssymb,nofootinbib]{revtex4}
\usepackage{amsmath}
\usepackage{dcolumn}
\usepackage{bm}
\usepackage[all, knot]{xy}
\xyoption{arc}
\begin{document}

\title{Cosmological Casimir effect with maximum planckian momentum and accelerating universe.}

\author{Fabio Briscese}
\email{briscese@dmmm.uniroma1.it}
\affiliation{Dipartimento di Modelli e Metodi Matematici and GNFM.\\
Universit\`a degli Studi di Roma "La Sapienza", Via A. Scarpa 16,
I-00161, Roma, Italy.}
\author{Antonino Marcian\`o}
\email{antonino.marciano@roma1.infn.it}
\affiliation{Dipartimento di Fisica and INFN\\
Universit\`a degi Studi di Roma "La Sapienza", P.le A. Moro 2,
00185 Roma, Italy}

\begin{abstract}
\begin{center}
{\bf Abstract}
\end{center}
We develop here a mechanism that, without making use of a
cosmological constant, reproduces an accelerating universe. This
is done by taking into account Casimir vacuum energy density,
assuming that the underlying theory allows a maximum momentum,
that turns out  to be the leading contribution term to Einstein
equations in a large expanding FRW universe. As stated in numerous
quantum gravity studies, we postulate that maximum momentum is
related to the existence of the Planck length as a fundamental
length. This insight, together with the assumption of a Planck
scale correction to the energy/momentum dispersion-relation on a
FRW background, is used here to calculate Casimir vacuum energy.
We show that, under these hypothesis, an accelerated universe
expansion is obtained. As last step we analyze the compatibility
of the resulting model with experimental data, writing down the
equation of state for Casimir energy and pressure and observing
that this equation of state belongs to a class of models that
naturally fits cosmological observations. We emphasize that our
result relies, once a fundamental length is introduced in Casimir
effect, just on general arguments thus it is independent on an
explicit form of the energy-momentum dispersion relation.
\end{abstract}
\maketitle

\section{Introduction.}
As evident by consolidated astrophysical data, the universe is in
accelerating expansion. To account for this fact, in the
$\Lambda$-CDM model a
cosmological constant has been introduced, with the meaning of Dark Energy (for a review see \cite{od, Ratra-Peebles, Cald} and related refs.), that counts for about $70\%$ of the total mass energy density. With the purpose of explaining the introduction of such a cosmological constant some authors have dealt with Casimir effect, but without making use of a maximum planckian momentum, see for exemple \cite{eli}.  An alternative to the standard  cosmological model is represented, for exemple, by modified gravity, as discussed in \cite{od2, od4, od5}, and by phantom models, \cite{od3, od6} and related refs.\\
Here we want to present a mechanism for the accelerating universe that relies upon two hypothesis: 1) that a maximum momentum exists and that it is related to a fundamental scale, the Planck length; 2) that a Casimir vacuum energy contributes to the energy density of the universe. \\
There has been a long debate in quantum gravity literature (for a satisfactory review see \cite{carlip}) about possible ways of introducing Planck length as a quantum gravity scale \cite{Garay, rova2}.\\
Despite of severe difficulties in obtaining experimental data in which the role of Planck scale manifestly appears, in recent years a wide and increasing literature on quantum gravity phenomenology has been produced. Possible connections to forthcoming experiments have been explained and suitable observable quantities sensible to planckian structure of space-time have been identified \cite{ame3, ame5}. As a consequence, Planck scale corrections to the energy momentum dispersion relation have been considered as the straightest way to approach these phenomenological issues. Planck scale can in fact appear by heuristic considerations, either as a deformation parameter of the special relativistic or general relativistic energy momentum dispersion relations, or as a maximum spatial momentum \cite{ame, mag2}. 
In this scenario, we first assume that analytical Planck scale
deformation to the one particle energy momentum dispersion
relation can be considered in a FRW expanding universe background.
\footnote{To be simple, we will restrict our attention only to
massless fields without any other internal degrees of freedom.
This assumption will not affect in a drastic way our results,
giving only small negligible corrections to the calculus shown
below or supplying multiplication for order one constant extra
factors. }
\\
Some authors refer to the Planck length as a minimum observable length \cite{carlo}, that is responsible for cutting off trans-planckian degrees of freedom. Specially in Loop Quantum Gravity, recent works \cite{Ashte, bojo} clarify how the very planckian structure of the space-time, emerging in a consistent quantum gravity model, may involve a qualitatively different evolution of the universe in the (planckian) early times. \\
Anyway, even without making any strong assumptions on the deep
planckian regime of a full quantum gravity theory, one can assume
that if such a quantum gravity theory exists, Planck departures
from standard classical gravitational relations can be also
postulated at low energies or large distances \cite{Lambda, ame,
ame2}. Such a reasoning applies to the details of the analysis we
are going to develop here below. In this letter we want to show
that the introduction of the Planck length (although presented in
a heuristic/phenomenological way, assuming Planck scale
corrections to the one-particle energy momentum dispersion
relation and hence to the Casimir vacuum energy density) may  also
induce considerable modification to the behavior of a large
expanding universe, \i.e. a universe whose metric conformal factor
$a$ satisfies ${\lambda_{Planck}}<<{a}$. In  section II, we in
fact assume that a cosmological Casimir energy, calculated as a
massless scalar field vacuum energy, contributes\footnote{Such an
assumption could be discussed in depth highlighting possible
connections between classical and quantum field theories
\cite{pa2}.} to the right hand side of the Einstein equation.
Vacuum energy is then calculated assuming Planck scale modified
energy momentum dispersion relation for a single massless particle
with a maximum planckian momentum. An analytical expansion in
${\lambda_{Planck}}/{a}$ to the Casimir energy density is then
performed. The reasonableness condition taken into account,
$\lambda_{Planck}<<a$, of course limits the validity of our
reasonings to post-planckian cosmological eras.
\\
By these hypothesis, in section III we obtain a mechanism that is
different from the mechanisms of the models cited above, that can
account for an accelerated expansion of the universe without
postulating the existence of a cosmological constant. As last
step, in section IV, we write down the equation of state for the
Casimir energy density and pressure and the explicit form of the
Casimir energy density. We then show that the model obtained
belongs to a class of models that naturally fits cosmological
observations. In section V, as closing remarks, we summarize the
obtained results.

\section{Cosmological Casimir effect with a maximum planckian momentum.}
To start with, let us consider an homogeneous space-time in a
comooving FRW coordinates system. We can assume that the metric
tensor is given by
\begin{equation}
ds^2 = - (c dt)^2 + a(t)^2 [ d\chi^2 + \Sigma^2 ( d \theta^2 +
Sin^2(\theta) d\phi^2 ]
\end{equation}
In this coordinates system, Einstein equations read
\begin{equation}
\left\{
\begin{array}{rl}

G_{tt} = 3 \left( \frac{\dot{a}}{ c a} \right)^2 + 3 \frac{k}{a^2} = \frac{8 \pi G}{c^3} T_{tt} \\
\\

G_{\mu \mu} = -  \frac{2\ddot{a}}{c^2 a} - \left( \frac{\dot{a}}{c
a} \right)^2 - \frac{k}{a^2} = \frac{8 \pi G}{c^3}  T_{\mu \mu}

\end{array}
\right.
\end{equation}
where $ \mu = \chi,\theta,\phi $ is the spatial index; $ T_{\mu
\mu}$ is the pressure of the system, $T_{tt}$ is its energy
density and $\frac{k}{a^2}$ is the spatial scalar curvature. As
said in the introduction, maximum planckian momentum is introduced
in the energy momentum dispersion relation, so this relation
results modified. Using this hypothesis we can write the energy
momentum relation for a massless scalar field in the form
\begin{equation}
E = \hbar  \; \omega(|\stackrel{\rightarrow}{k}|,\lambda,a)
\end{equation}
We are considering for simplicity the case of a massless scalar
field, but the following argument can be easily generalized to
other kind of fields. The Casimir energy is given by

\begin{equation}
T_{tt} = \frac{E_0}{c V} = \hbar \int \frac{d^3k}{c(2\pi)^3}
\omega(k,\lambda,a) = \frac{2\hbar}{c(2\pi)^2}
\int^{\frac{1}{\lambda}}_{\frac{1}{2a}, 0} dk k^2
\omega(k,\lambda,a) = \frac {\hbar}{2\pi^2} F(\lambda , a)
\end{equation}
where $$ F(\lambda,a) = \frac{1}{c}
\int^{\frac{1}{\lambda}}_{\frac{1}{2 a}, 0} dk k^2
\omega(k,\lambda,a)
$$
In this equation, the integration starts from 0 for the open or
spatially flat FRW universe, from $\frac{1}{2 a}$ for the closed
FRW universe. Note that we are completely disregarding the
contribution of matter and radiation that, as it will be evident
later, is negligible in a large expanding universe. We then obtain
for the system of Einstein equations
\begin{equation}
\left\{
\begin{array}{rl}

\left( \frac{\dot{a}}{c a} \right)^2 + 3 \frac{k}{a^2} = \frac{4\hbar}{3\pi c^3} \, G \, F(\lambda,a)\\
\\

( T_{tt}a^3 ),_t = - T_{\mu \mu} (a^3),_t

\end{array}
\right.
\end{equation}
Here we substituted the  second  equation of (2) with the energy
conservation equation. The second equation in (5) is used just to
determine $T_{\mu \mu}$, so at the moment we can ignore it. Let us
now consider the function $F(\lambda,a)$. By dimensional analysis
it follows that
\begin{equation}
F(\lambda,a) = \frac{1}{\lambda^3}\left(
\frac{\alpha(\frac{\lambda}{a}) }{\lambda} +
\frac{\beta(\frac{\lambda}{a}) }{a}\right)
\end{equation}
In fact we can write
$$F(\lambda,a) = \frac{1}{c \lambda^3} \int^{1}_{\frac{\lambda}{2 a}\; ,\; 0} dx \;x^2\; \omega(\frac{x}{\lambda},\lambda,a)$$
Note that $\frac{\omega}{c} $ has the dimension of an inverse of
length so that the only way to write it is
$$ \frac{\omega}{c} =  \sum_{k=0}^\infty \left( \frac{A_k (x) }{\lambda} + \frac{B_k (x) }{a}\right) \left(\frac{\lambda}{a}\right)^k $$
thus we have
$$F(\lambda,a) = \frac{1}{c \lambda^3}\int^{1}_{\frac{\lambda}{2 a}\; ,\; 0} dx \;x^2\; \omega(\frac{x}{\lambda},\lambda,a)=
\frac{1}{\lambda^3} \sum_{k=0}^\infty \left( \frac{A_k  }{\lambda}
+ \frac{B_k  }{a}\right) \left(\frac{\lambda}{a}\right)^k =
\frac{1}{\lambda^3} \left( \frac{\alpha(\frac{\lambda}{a})
}{\lambda} + \frac{\beta(\frac{\lambda}{a}) }{a}\right)
$$
Here, $\alpha $ and $\beta$ are analytic functions in
$\frac{\lambda}{a}$.
$$
\left\{
\begin{array}{rl}

\alpha(\frac{\lambda}{a}) = \sum_{k=0}^\infty \alpha_k (    \frac{\lambda}{a} )^k \\

\beta(\frac{\lambda}{a}) = \sum_{k=0}^\infty \beta_k (
\frac{\lambda}{a} )^k
\end{array}
\right.
$$
In order to obtain  the net Casimir energy $F_{net}(\lambda,a)$,
we have to subtract to this quantity its infinite limit
\begin{equation}
F(\lambda, \infty) = lim_{a \rightarrow \infty } F(\lambda,a) =
\frac{\alpha(0)}{\lambda^4}
\end{equation}
so that
\begin{equation}
F_{net}(\lambda,a) = \frac{1}{\lambda^3} \left(
\frac{\alpha(\frac{\lambda}{a})
 - \alpha(0)}{\lambda} + \frac{\beta(\frac{\lambda}{a}) }{a}\right) = \frac{1}{\lambda^3 a} B(\frac{\lambda}{a})
\end{equation}
in which $B$ is an analytic function of $\lambda / a $ with $ B(0)
\neq 0 $. This is our final expression for $F(\lambda,a)$.

\section{The accelerating universe.}
Now we can go back to the first Einstein equation and write it in
the form:
\begin{equation}
\left(\frac{\dot{a}}{c}\right)^2 = - k + \frac{4 \hbar G}{3 \pi
c^3} \frac{a}{\lambda^3} B(\frac{\lambda}{a})
\end{equation}

It is evident  that this is an equation for an accelerating
universe. Now we can set set $\lambda = \lambda_{Planck}$ and
obtain
\begin{equation}
\left(\frac{\dot{a}}{c}\right)^2 = - k + \frac{4 }{3 \pi }
\frac{a}{\lambda_{Planck}} B(\frac{\lambda_{Planck}}{a})
\end{equation}
For the leading term we find
\begin{equation}
\left(\frac{\dot{a}}{c}\right)^2 = - k + \frac{4 }{3 \pi }
\frac{a}{\lambda_{Planck}} B(0)
\end{equation}
Note that matter and radiation densities are completely negligible
in a large universe, because of the fact that they are
respectively of order $\frac{1}{a^3}$ and $\frac{1}{a^4}$. We want
to stress that in order to write this relation we used (3).
Although we do not know explicitly (3) and hence the
corresponding\footnote{As an example, one may consider, without
any particular physical intent, the case of the following
energy/momentum dispersion relation $\omega(k,\lambda,a) =
\frac{c}{\lambda} \ln \left( \frac{1}{1- \lambda k }  \right)
\left( 1 + \frac{\lambda}{a} \right)$, from wich follows that
$B(0) = \frac{11}{18}$} $F(\lambda,a)$, we are able to predict the
universe accelerating expansion in the limit of large $a(t)$. We
also stress that this discussion is based on dimensional analysis,
so it is, after the introduction of the fundamental Planck length,
totally general. Now we can ask if relation (11) agree with
cosmological data. To answer to this question we have to write the
equation of state for Casimir Energy and Pressure and write the
Casimir energy density as a function of $a(t)$.

\section{Equation of state and cosmological observations.}
To obtain equation of state for the Casimir Energy we have to use
the second of equations in (5). Using (4) and (8) we have
\begin{equation}
T_{tt} = \frac{\hbar}{2 \pi^2} \frac{B(\frac{\lambda}{a})}{a
\lambda^3}
 \simeq \frac{\hbar}{2 \pi^2} \frac{B(0)}{a \lambda^3}
\end{equation}
In this approximation we have
\begin{equation}
T_{\mu \mu} = - \frac{2}{3} T_{tt}
\end{equation}
We note that this result is in agreement with experimental data.
In fact, as discussed in \cite{Cald}, equation of state with $-1 <
\omega < 0$, where $\omega$ is the ratio of the pressure to the
energy density, fits current cosmological observation best. So, as
follows from the last equation, in our case we have $\omega = -
\frac{2}{3}$, and this value belongs to the range mentioned above.
Moreover, in order to confront our model with experimental data,
we can link our parameter $B(0)$ with $a_0$ and $H_0$,
respectively the scale factor today and the Hubble constant today.
From (9), setting the spatial curvature equal to zero, in
agreement with WMAP observations \cite{WMAP} \cite{paolo}, we have

\begin{equation}
\left\{
\begin{array}{rl}
\rho_{casimir} = B(0)  \rho_c \frac{a_0}{a}  \\
\\
a_0 = \frac{4 c^2}{3 \pi \lambda_{Planck} \; H_o^2 }\\
\\
\rho_c = \frac{3 \hbar H_0^2}{8 \pi c \lambda_{Planck}^2}
\end{array}
\right.
\end{equation}
where $\rho_c$ is the critical energy density. It is evident that
$B(0)$ simply represent the ratio between the Casimir energy
density and the critical density. Note that $B(0)$ is a pure
number thus it would be desirable for it , following a naturalness
criterion, to take values in the neighborhood of the unity. This
is also in agreement with cosmological observations, that predict
a value for this parameter close to $0.6-0.7$ \cite{Cald}. In
light of these facts, we conclude that our model is a good
candidate to explain the accelerating expansion of the universe.

\section{Conclusions.}
We conclude this letter remarking the fundamental points of our
analysis. We first used the hypothesis of the existence of a
maximum momentum related to the Planck scale and we calculated the
Casimir energy density of a FRW expanding universe. This mechanism
actually reproduces an accelerating universe. We want to emphasize
that this result follows from dimensional analysis. At the end, we
obtained the equation of state for Casimir energy and pressure and
the expression of Casimir energy density as a function of the
scale factor. These expressions are in agreement with current
cosmological data. A further analysis is needed to study the
compatibility of this toy model with CMB observations. In
conclusion, this toy model can offer a mechanism to explain the
accelerating expansion of the universe and it can be easily
improved to give a real physical model, without affecting the
fundamental result, by the inclusion of dark matter and other
contributions to the total energy density.

\section*{Acknowledgements.}
We are very grateful to Giovanni Amelino-Camelia for useful
discussions during the developing of this study. We want also to
thank Paolo Serra for useful discussions on the cosmological
observations, especially the ones resulting from WMAP.


\addcontentsline{toc}{chapter}{Bibliography}
\begin{thebibliography}{9}

\bibitem{od} T.Padmanabhan {\it Cosmological constant: The Weight of the vacuum.} published in Phys. Rept. 380:235-320, 2003; hep-th/0212290
\bibitem {Ratra-Peebles} P.J.E. Peebles, B. Ratra {\it The Cosmological constant and dark energy.} published in Rev.Mod.Phys. 75:559-606, 2003; astro-ph/0207347
\bibitem{Cald} R.R.Caldwell, Rahul Dave and Paul J. Steinhardt. Cosmological Imprint of an Energy Component with General Equation-of-State. Published in  Physical Review Letters 80 1582 (1998). http://arxiv.org/abs/astro-ph/9708069.
\bibitem{eli} E. Elizalde {\it Cosmological uses of Casimir energy.} published in *Norman 2003, Quantum field theory under the influence of external conditions* 317-324; hep-th/0311195
\bibitem{od2} S. Nojiri and S.D.Odintsov. {\it Introduction to  modified gravity and gravitational alternative for dark energy} published in Int.J.Geom.Meth.Mod.Phys.4: 115-146, 2007; hep-th/0601213
\bibitem{od4}
 S. Nojiri, S. D. Odintsov {\it Modified gravity as an alternative for Lambda-CDM cosmology.} presented at 2nd International Conference on Quantum Theories and Renormalization Group in Gravity and Cosmology (IRGAC 2006), Barcelona, Spain, 11-15 Jul 2006; hep-th/0610164
\bibitem{od5} S. Capozziello, S. Nojiri, S.D. Odintsov {\it  Dark energy: The Equation of state description versus scalar-tensor or
modified gravity.} published in Phys.Lett.B634: 93-100, 2006;
hep-th/0512118
\bibitem{od3} F. Briscese, E. Elizalde, S. Nojiri, S.D. Odintsov
{\it Phantom scalar dark energy as modified gravity: Understanding
the origin of the Big Rip singularity.} published in
Phys.Lett.B646: 105-111, 2007; hep-th/0612220
\bibitem{od6}
 S. Capozziello, S. Nojiri, S.D. Odintsov {\it Unified phantom cosmology: Inflation, dark energy and dark matter under
the same standard.} published in Phys.Lett.B632: 597-604, 2006;
hep-th/0507182
\bibitem{Garay} L.~J.~Garay {\it Quantum gravity and minimum length.}
  Int.\ J.\ Mod.\ Phys.\ A {\bf 10}, 145, 1995; gr-qc/9403008
\bibitem{rova2} C. Rovelli {Notes for a brief history of quantum gravity.} Published in *Rome 2000, Recent developments in theoretical and experimental general relativity, gravitation and relativistic field theories, Pt. A* 742-768; gr-qc/0006061
\bibitem{carlip}
S. Carlip {\it Quantum gravity: A Progress report.} UCD-2001-04,
Mar 2001; published in Rept. Prog. Phys. 64: 885, 2001;
gr-qc/0108040
\bibitem{ame3}
G. Amelino-Camelia, J. R. Ellis, N.E. Mavromatos, D. V.
Nanopoulos, S. Sarkar {\it Tests of quantum gravity from
observations of gamma-ray bursts.} Published in Nature 393:
763-765,1998; astro-ph/9712103
\bibitem{ame5} G. Amelino-Camelia
{\it Proposal of a second generation of quantum gravity motivated
Lorentz symmetry tests: Sensitivity to effects suppressed
quadratically by the Planck scale.} Published in
Int.J.Mod.Phys.D12:1633-1640, 2003; gr-qc/0305057
\bibitem{carlo}
C. Rovelli, Lee Smolin {\it Discreteness of area and volume in
quantum gravity.} CGPG-94-11-1, Nov 1994. published in Nucl.Phys.
B442:593-622, 1995, Erratum-ibid.B456:753,1995; gr-qc/9411005
\bibitem{ame} G. Amelino-Camelia {\it Doubly special relativity.} Published in Nature 418: 34-35, 2002; gr-qc/0207049
\bibitem{mag2} J. Magueijo, L. Smolin {\it Generalized Lorentz invariance with an invariant energy scale.} Published in Phys.Rev.D67: 044017, 2003; gr-qc/0207085
\bibitem{Ashte} A. Ashtekar, M. Bojowald, J. Lewandowski {\it Mathematical structure of loop quantum cosmology.} Published in Adv.Theor.Math.Phys.7:233-268, 2003; gr-qc/0304074
\bibitem{bojo} M. Bojowald {\it Loop quantum cosmology.} Published in Living Rev.Rel.8:11, 2005; gr-qc/0601085
\bibitem{pa2} K. Srinivasan, L. Sriramkumar, T. Padmanabhan {\it Plane waves viewed from an accelerated frame: Quantum physics in classical setting.} Published in Phys.Rev.D56: 6692-6694,1997;
\bibitem{Lambda}
G. Amelino-Camelia, L. Smolin, A. Starodubtsev, {\it Quantum
symmetry, the cosmological constant and Planck scale
phenomenology.} Published in Class.Quant.Grav.21: 3095-3110, 2004;
hep-th/0306134
\bibitem{ame2} G. Amelino-Camelia {\it  Relativity in space-times with short distance structure governed by an observer independent (Planckian) length scale.} Published in Int.J.Mod.Phys.D11: 35-60, 2002; gr-qc/0012051
\bibitem{WMAP} WMAP Collaboration (D.N. Spergel et al.) {\it Wilkinson Microwave Anisotropy Probe (WMAP) three year results: implications for cosmology.} Submitted to Astrophys.J. astro-ph/0603449
\bibitem{paolo} P. Serra, A. Heavens, A. Melchiorri {\it Bayesian evidence for a cosmological constant using new high-redshift Supernovae data.} astro-ph/0701338
\end{thebibliography}
\end{document}